\documentclass[amsmath,amssymb,prl,aps,10pt,superscriptaddress,twocolumn]{revtex4-2}
\usepackage{graphicx,xcolor,times}
\begin{document}
\title{Theory of distribution skewness effect on polydisperse random close packing}

\author{Vinay Vaibhav}
\email{vinayphys@gmail.com}
\affiliation{Department of Physics ``A. Pontremoli'', University of Milan, via Celoria 16, 20133 Milan, Italy.}

\author{Carmine Anzivino}
\email{carmine.anzivino@gmail.com} 
\affiliation{Department of Physics ``A. Pontremoli'', University of Milan, via Celoria 16, 20133 Milan, Italy.}

\author{Alessio Zaccone}
\email{alessio.zaccone@unimi.it}
\affiliation{Department of Physics ``A. Pontremoli'', University of Milan, via Celoria 16, 20133 Milan, Italy.}

\date{\today}
 
\begin{abstract}
We investigate the random close packing density, $\phi_\textrm{RCP}$, of polydisperse hard sphere systems using a theoretical framework based on the equilibrium model of crowding. We derive a closed-form solution for $\phi_\textrm{RCP}$ in terms of the moments of the diameter distribution, enabling an analytical exploration of the effects of polydispersity ($\delta$) and skewness ($S$) on packing density. For a binary mixture, it is possible to explore a broader range of dependence of $\phi_\textrm{RCP}$ on $\delta$ for a given $S$ or on $S$ for a given $\delta$. We show that the dependencies of $\phi_\textrm{RCP}$ on skewness for a variety of continuous distributions collapse onto a theoretical master curve obtained for the binary mixture case. By correcting the theory so that it obeys known exact limiting behaviours for extreme size asymmetry, our analytical predictions not only agree with previously obtained numerical results, but also predict previously unexplored regions of the $\phi_\textrm{RCP}$ parameter space.
\end{abstract}

\maketitle

Hard-sphere systems serve as a fundamental and extensively studied model in statistical physics and condensed matter, with applications ranging from granular media to colloidal suspensions \cite{hansen2013theory, torquato2010jammed, royall2024colloidal}. Since hard spheres interact only through excluded volume, their equilibrium structure and phase behavior are controlled by the packing fraction $\phi$, defined as the fraction of volume occupied by the spheres, rather than by temperature. Upon slow compression, a monodisperse hard-sphere fluid undergoes an entropy-driven first-order freezing transition, with fluid–solid coexistence between $\phi_\textrm{freeze} \approx 0.494$ and $\phi_\textrm{melt} \approx 0.545$. Above coexistence, the stable crystalline phase can reach the face-centered cubic close-packed limit  $\phi_\textrm{fcc} \approx 0.740$ \cite{alder1957phase, wood1957preliminary, hoover1968melting, torquato2010jammed}. However, crystallization can be suppressed, especially under rapid compression, leading to jammed, disordered configurations at lower volume fractions characterized by the random close packing (RCP) density $\phi_{\rm RCP}$ \cite{bernal1960packing, o2003jamming, sanz2011crystallization}. Although the precise definition of RCP remains debated \cite{torquato2010jammed, kamien2007random}, it is generally associated with the maximally random jammed (MRJ) state \cite{torquato2000random,PhysRevE.62.993}, where the system is isostatic and the average contact number matches the mechanical stability limit ($z=6$ in three dimensions for frictionless spheres, excluding rattlers) \cite{zaccone2022explicit}.

Polydispersity, the presence of particles with a distribution of sizes, plays a critical role in the packing behavior. In polydisperse HS mixtures, crystallization is significantly suppressed, and the denser packing configurations can arise because smaller spheres partially fill the voids between larger particles, increasing the achievable $\phi_{\rm RCP}$ beyond the monodisperse limit \cite{donev2006binary, desmond2014influence, anzivino2023estimating, meer2024bounded}. Owing to its relevance for applications, the RCP of polydisperse hard spheres has attracted a great deal of attention in recent years and has been widely investigated by means of theory, simulations, and experiments \cite{mcgeary1961mechanical, phan1998effects, castillo2008solving, jerkins2008onset, dorr2013discrete, danaei2018impact, C000984A, desmond2009random,  vaibhav2022finite, biazzo2009theory, vaibhav2022rheological, jiang2023effects}. While polydispersity (defined as the standard deviation divided by the mean of diameters) is essential, it does not fully characterize the influence of the size distribution on packing. Higher moments, most notably the skewness $S$ of the diameter distribution, have a significant effect on $\phi_{\rm RCP}$, whereas higher moments such as kurtosis appear to have a weaker influence in many studied cases \cite{desmond2014influence, meer2024bounded}. The pioneering work by Desmond and Weeks \cite{desmond2014influence} systematically demonstrated through simulations that both polydispersity (or coefficient of variation) $\delta$ and skewness $S$ control the RCP density. They proposed an empirical relation and showed, at fixed $\delta$, $\phi_{\rm RCP}$ increases approximately linearly with $S$, regardless of the specific form of the size distribution. More recent studies \cite{meer2024estimating} have refined this picture, showing that the linear dependence breaks down near limiting values of $S$ corresponding to extreme size asymmetry. From a theoretical perspective, Santos \emph{et al.} \cite{santos2014PRE, santos2020JCP, santos2026predicting} introduced a compact, moment-based approximation that encodes the effect of polydispersity and skewness into a single parameter,
$\lambda = (1 + 3\delta^2 + S \delta^3)/{(1 + \delta^2)^2}$,
which predicts that all mixtures sharing the same $\lambda$ exhibit similar $\phi_{\rm RCP}$ values. Despite these advances, a general analytical theory capable of describing the combined effects of $\delta$ and $S$, while simultaneously satisfying the exact asymptotic limits of highly asymmetric mixtures, is lacking.

In this study, we critically evaluate the role of polydispersity and skewness in determining $\phi_{\rm RCP}$ and propose a general theoretical framework for the random close packing of polydisperse hard spheres. Starting from a truncatable three-moment description of the mixture equation of state, we derive a closed-form prediction for $\phi_{\rm RCP}(\delta,S)$. Within this approximation, all size distributions with the same polydispersity $\delta$ and skewness $S$ are predicted to have the same $\phi_{\rm RCP}$, allowing continuous distributions to be mapped onto an effective bidisperse system with matching moments. We explicitly verify this mapping for several families of size distributions $P(\sigma)$. We further modify the theory to satisfy the exact asymptotic limits associated with extreme size asymmetry and vanishing-size rattlers. The resulting predictions are consistent with available numerical simulations and experimental data, while also extending into previously unexplored regions of the $\phi_{\rm RCP}$ parameter space. These results clarify the interplay between polydispersity and skewness in controlling random close packing and provide a basis for further theoretical and applied studies.

 \begin{figure*}
    \centering
    \includegraphics[width=8.75cm]{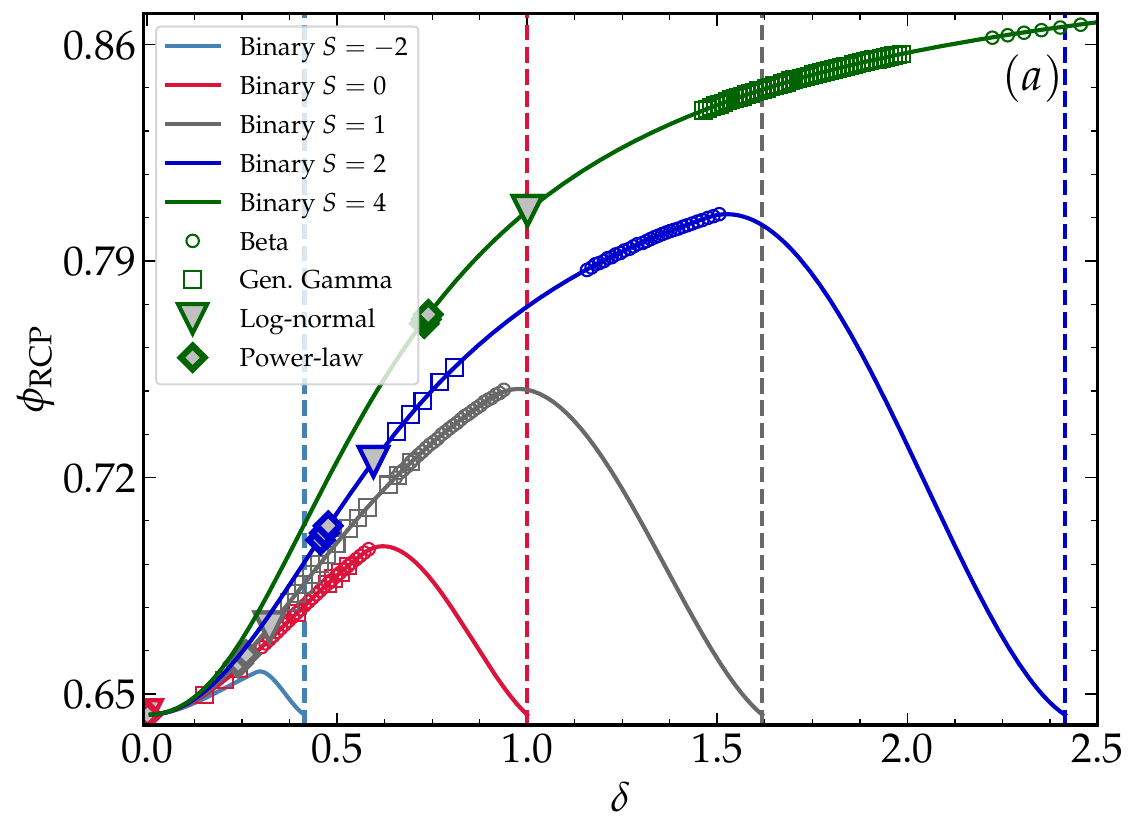}
    \includegraphics[width=8.75cm]{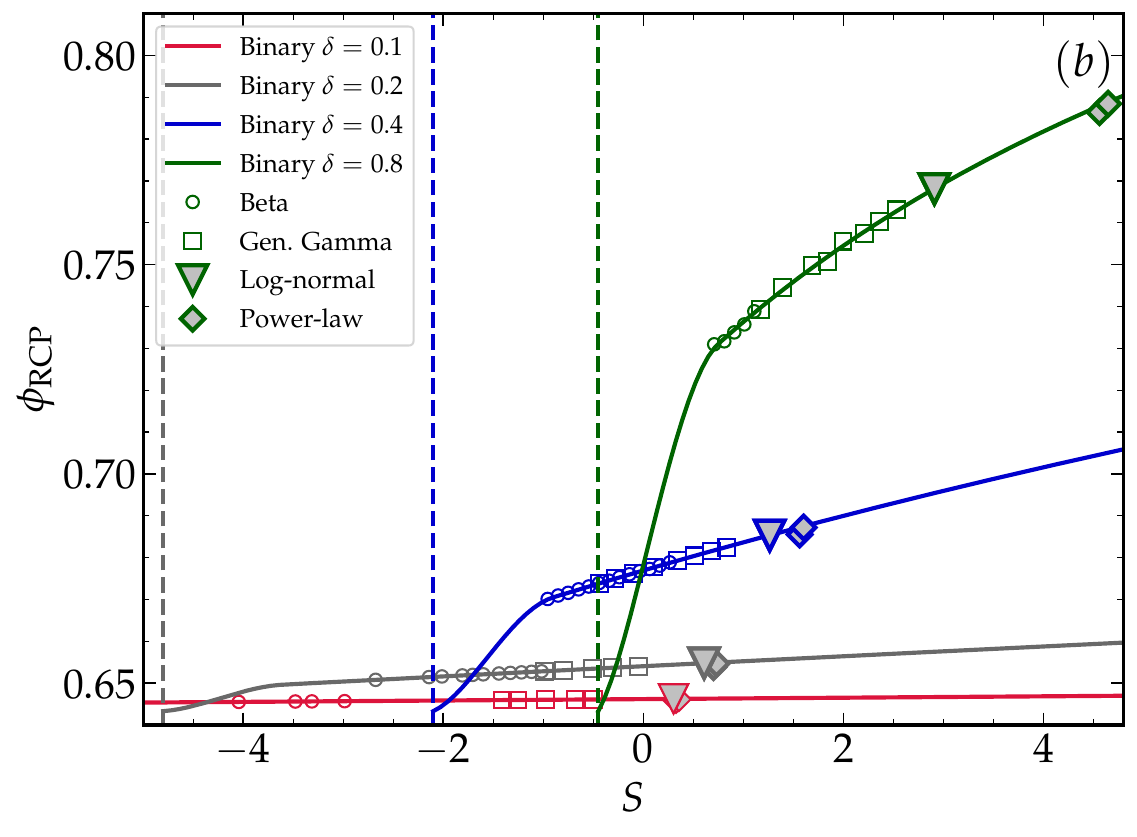}
    \caption{{ Variation of random close packing fraction with polydispersity and skewness.} $\phi_{\rm RCP}$ as a function of (a) polydispersity $\delta$ for different values of skewness $S$, as indicated, and (b) as a function of skewness $S$ for different values of polydispersity $\delta$, as indicated. Solid lines correspond to the analytical prediction $\phi_{\rm RCP}(\delta,S)$ obtained from the three–moment theory (see text and Supplementary Information~\cite{supp}), while symbols denote the same function evaluated for different continuous size distributions: beta (open circles), generalized gamma (open squares), log-normal (closed triangles), and truncated power-law (closed diamonds). Symbols color correspond to the case of fixed $S$ in (a) and fixed $\delta$ in (b) as indicated with the same color for the binary distribution. Dashed vertical lines in panel (a) mark the upper bound $\delta_{\rm max} = (S+M)/2$ for given $S$, while those in panel (b) indicate the lower bound $S_0 = \delta - (1/\delta)$ for given $\delta$ (see text for details).}
    \label{fig:deltaS}
\end{figure*}

The recent analytical theory for RCP of monodisperse spheres \cite{zaccone2022explicit} has opened up the possibility of having an analytical prediction for the RCP of hard spheres with size disparity, with the diameters of spheres following some size distribution \cite{anzivino2023estimating}. Such analytical predictions are in agreement with the numerical simulations, as shown in \cite{anzivino2023estimating}. The key insights of this novel approach are that the RCP can be identified with the densest MRJ state that is marginally rigid ($z=6$), and that a reasonable model of crowding for MRJ states can be obtained on the basis of approximate liquid theories.

We consider an $m$-component mixture with the number of spheres of type $i$ of diameter $\sigma_{i}$ denoted as $N_i$, such that  $\sum_{i=1}^{m} N_i = N$ is the total number of particles in the system, and the corresponding number fraction of species $i$ is $x_i = N_i/N$. Also, the contact distance between a sphere of species $i$ and a sphere of species $j$ can be denoted as $\sigma_{ij} = \frac{1}{2} (\sigma_{i} + \sigma_{j})$. The $n$th moment of diameter can be defined as $\left\langle \sigma^{n} \right\rangle \equiv \sum_{i=1}^{m} x_{i} \sigma_{i}^{n}$. In this case, the volume fraction of the system is $\phi = \pi \rho \left\langle \sigma^{3} \right\rangle /6$, where $\rho$ is the number density. The coordination number $z_{ij}$ between particles of species $i$ and $j$ can be written in terms of the partial pair-correlation function $g_{ij}$ \cite{hansen2013theory, mulero2008theory}:
\begin{equation}\label{int}
    z_{ij} = 4 \pi \rho \int_{0}^{\sigma_{ij}^+} dr~r^2 g_{ij}(r),
\end{equation}
where $r$ is the inter-particle radial separation \cite{anzivino2023estimating}. In the above expression, $\sigma_{ij}^+ = \sigma_{ij} + \epsilon$ where $\epsilon$ is in an infinitesimally small positive number. Following the key idea proposed in Ref. \cite{zaccone2022explicit}, we write the pair-correlation $g_{ij}(r)$ as a partially continuous function \cite{torquato2018perspective}, i.e., $g_{ij}(r) = g_{ij,{\rm C}}(\sigma_{ij};\phi)\delta(r-\sigma_{ij}) + g_{ij,{\rm BC}}(r)$, where the first term represents the discrete part, i.e. the probability of nearest neighbors in direct contact with the central particle and the second term represents the continuous part, i.e. the probability of finding the particles beyond contact. Using this idea if we evaluate the integral in Eq.~(\ref{int}), the only part of $g_{ij}(r)$ which contributes to the coordination number is the contact value $g_{ij,{\rm C}}$. If we choose the ansatz \cite{anzivino2023estimating, ansatz} $g_{ij,{\rm C}}(\sigma_{ij}; \phi) \equiv \frac{\sigma_{ij}}{\langle \sigma \rangle} g_0(\langle \sigma \rangle) g_{ij}(\sigma_{ij}; \phi),$ where $g_0$ is a dimensional constant (with dimension $1/length$), the following expression for the species-averaged coordination number $z$ can be obtained \cite{anzivino2023estimating}:
\begin{align}
    \langle z \rangle  &= 24 \phi \frac{g_0}{\langle\sigma\rangle} \sum\limits_{i,j}^{m} x_i x_j \frac{\sigma_{ij}^3}{\langle \sigma^3 \rangle} g_{ij}(\sigma_{ij}; \phi).\label{key}
\end{align}

It is known that for an $m$-component hard sphere fluid in equilibrium, the equation of state (EOS) can be written in terms of the compressibility factor $\mathcal{Z}^{(m)}$ \cite{lebowitz1964exact,mansoori1971equilibrium,santos1999equation}:
\begin{equation} \label{Z_mixture}
    \frac{\mathcal{Z}^{(m)} (\phi) - 1}{4 \phi} =  \sum_{i=1}^{m} \sum_{j=1}^{m} x_{i} x_{j} \frac{\sigma_{ij}^{3}}{\left\langle \sigma^{3} \right\rangle} g_{ij}^{\rm eq} (\sigma_{ij}; \phi).
\end{equation}
Assuming that the local crowding in the MRJ state is similar to the corresponding equilibrium state, then the local structure near contact in the jammed state can be approximated by the structure near contact in the equilibrium fluid, i.e. $g_{ij}(\sigma_{ij}; \phi) \approx g_{ij}^{\rm eq}(\sigma_{ij}; \phi)$. Such an assumption has been justified extensively via numerical data in \cite{anzivino2023estimating}. In light of such an assumption, comparing Eqs.~(\ref{key}) and (\ref{Z_mixture}) yields, $6 C_0 (\mathcal{Z}^{(m)} (\phi) - 1)  = \langle z \rangle$, where $\langle\sigma\rangle/g_0 \equiv C_0$, to be determined based on a limiting reference state (usually taken as the closest packing of spheres in 3D provided by the face-centered cubic \cite{hales2005proof} or body-centered cubic \cite{likos} lattice). The condition of isostaticity i.e. $\langle z \rangle = 6$ is always satisfied at RCP, for both monodisperse and polydisperse systems. Comparing the generalized case of an $m$-component system at RCP with the monodisperse case at RCP, $6 C_0 (\mathcal{Z}^{(m)} (\phi_{\rm RCP}) - 1) = 6 = 6 C_0 (\mathcal{Z}^{(1)} (\phi_{\rm RCP}^{\rm M}) - 1)$, where $\phi_{\rm RCP}^{\rm M}$ is the monodisperse random packing fraction. This further simplifies to $\mathcal{Z}^{(m)} (\phi_{\rm RCP}) = \mathcal{Z}^{(1)} (\phi_{\rm RCP}^{\rm M})$. Within the present approach we now adopt a \emph{truncatable} representation of the mixture EOS, in which $\mathcal{Z}^{(m)}(\phi)$ is expressed in terms of $\phi$ and the first few moments of $P(\sigma)$ only. In particular, we use an extension of the Percus–Yevick (PY) virial equation of state for mixtures, which depends on the first three moments $\langle\sigma\rangle$, $\langle\sigma^2\rangle$ and $\langle\sigma^3\rangle$ \cite{santos1999equation}. This implies that $\phi_{\rm RCP}$ depends on $P(\sigma)$ only through the dimensionless combinations $\delta$ and $S$ defined below. More accurate non–truncatable EOS (e.g. PY–compressibility, BMCSL) could be used instead; they lead to cubic equations for $\phi_{\rm RCP}$ and slightly improved quantitative accuracy, but at the cost of losing an explicit closed form. A comparison with such alternatives is presented in the Supplementary Information \cite{supp}.

If we consider the Percus-Yevick (PY) equation of state (EOS) for the monodisperse system then $\mathcal{Z}^{(1)} (\phi_{\rm RCP}^{\rm M}) = (1+2\phi_{\rm RCP}^{\rm M}+3(\phi_{\rm RCP}^{\rm M})^2)/(1-\phi_{\rm RCP}^{\rm M})^2$, where $\phi_{\rm RCP}^{\rm M} \approx 0.643320$ if the body-centered cubic boundary condition is considered \cite{likos}. Assuming the generalized PY EOS for the $m$-component polydisperse mixture, the compressibility factor $\mathcal{Z}^{(m)} (\phi)$ is given in terms of the moments of the diameter distribution \cite{santos1999equation}
\begin{equation} \label{Zsantos}
\begin{aligned}
    \mathcal{Z}^{(m)}_{PY} & (\phi) = 1 + \big[ \mathcal{Z}_{PY} (\phi) - 1 \big] \frac{\left\langle \sigma^{2} \right\rangle}{2 \left\langle \sigma^{3} \right\rangle^{2}} \big( \left\langle \sigma^{2} \right\rangle^{2} + \left\langle \sigma \right\rangle \left\langle \sigma^{3} \right\rangle \big) \\
    &+ \frac{\phi}{(1-\phi)} \bigg[ 1 - \frac{\left\langle \sigma^{2} \right\rangle}{\left\langle \sigma^{3} \right\rangle^{2}}  \big( 2 \left\langle \sigma^{2} \right\rangle^{2} - \left\langle \sigma \right\rangle \left\langle \sigma^{3} \right\rangle \big) \bigg],
\end{aligned}
\end{equation}
where $\mathcal{Z}_{PY} (\phi) = (1+2\phi+3\phi^2)/(1-\phi)^2$ is the monodisperse PY compressibility factor \cite{hansen2013theory}. Hence, to find the RCP of a polydisperse system we first solve the PY–based matching condition
\begin{equation}\label{polydisSol}
  \mathcal{Z}^{(m)}_{PY}(\phi_{\rm RCP}) = \mathcal{Z}^{(1)} (\phi_{\rm RCP}^{\rm M}), 
\end{equation}
which yields a closed–form prediction $\phi_{\rm RCP}^{(0)}(\delta,S)$ (see Supplementary Information \cite{supp}). In a second step, discussed in detail in the Supplementary Information, we correct $\phi_{\rm RCP}^{(0)}(\delta,S)$ so as to enforce two exact limiting behaviours known for extreme size asymmetry (small–sphere and $w$–fixed limits), which the simple matching condition applied to the total pressure does not satisfy. The resulting corrected prediction is denoted by $\phi_{\rm RCP}(\delta,S)$ below. It should be noted that this approach is equally valid for a continuously polydisperse system $m\to \infty$. As per Eq.~(\ref{Zsantos}), only the first three moments of the diameter distribution are required as input.

We define the polydispersity $\delta$ and skewness $S$ of the probability distribution function of diameters $P(\sigma)$ as $\delta = \sqrt{\langle \Delta\sigma^2 \rangle}/\langle \sigma \rangle$ and $S = \langle \Delta\sigma^3 \rangle/\langle \Delta\sigma^2 \rangle^{3/2}$, respectively. Here, $\Delta\sigma = (\sigma - \langle \sigma \rangle)$, and the moments are given by $\langle \sigma^n \rangle = \int \sigma^nP(\sigma)~d\sigma$ and $\langle \Delta\sigma^n \rangle = \int \Delta\sigma^nP(\sigma)~d\sigma$ for $n \ge 0$. A distribution function with a high value of polydispersity corresponds to a high probability of occurrence of relatively large and small particles with respect to the mean diameter, and vice-versa. However, if the skewness of the distribution is large, then the probability of large diameters is very small, while small skewness signifies a low probability of occurrence of small diameters.

At first we consider the case of the discrete size distribution of binary mixtures, with the diameters of two species $\sigma_1$ and $\sigma_2$, for which the size distribution is $P(\sigma) = x_1\delta(\sigma-\sigma_1) + x_2\delta(\sigma-\sigma_2)$ and the $n$th moment of the diameter distribution is $\langle \sigma^n \rangle = x_1\sigma_1^n + x_2\sigma_2^n$. Coefficient of variation $\delta$ and skewness $S$ can be calculated as $\delta = (\eta-1)\sqrt{x_2-x_2^2}/(1-x_2+\eta x_2)$ and $S = (1-2x_2)/\sqrt{x_2-x_2^2}$, where $\eta = \sigma_2/\sigma_1 > 1$. These can be inverted to obtain $\eta$ and $x_2$ in terms of $\delta$ and $S$: $\eta = (2+\delta(S+M))/(2+\delta(S-M))$ and $x_2 = (M-S)/(2M)$, where $M = \sqrt{4+S^2}$ \cite{meer2024bounded}. Now, Eq.~(\ref{polydisSol}) can be solved for $\phi_{\rm RCP}$ using the moments in terms of $\eta$ and $x_2$, which can be eliminated in terms of $\delta$ and $S$ to obtain $\phi_{\rm RCP}(\delta, S)$ (see Supplementary Information \cite{supp} for the analytical form). These results are plotted as solid lines in Fig.~\ref{fig:deltaS} as functions of $\delta$ for different values of $S$ and as functions of $S$ for different values of $\delta$. The $\phi_{\rm RCP}$ initially increases with increasing polydispersity $\delta$ at fixed skewness S, reaches a maximum, and then bends downward toward the monodisperse limit $\phi_{\rm RCP}^{M}$ as $\delta \to \delta_{\max}$. This behavior follows from the exact asymptotic constraint associated with vanishing-size rattlers and extreme size asymmetry. Similarly, at fixed $\delta$, all curves $\phi_{\rm RCP}(S)$ originate from the common value $\phi_{\rm RCP}^{M}$ at the cutoff $S=S_0$. The dependence of $\phi_{\rm RCP}$ on $S$ at fixed $\delta$ is generally increasing over the physically accessible range. However, the behavior for small $\delta$ is roughly linear for a reasonably broad range of $S$, and with the increase of $\delta$ it eventually becomes non-linear for the whole range of $S$. At sufficiently large values of $S$, the increase of $\phi_{\rm RCP}(S)$ progressively weakens due to the geometric upper bound imposed by the hard-sphere packing constraint. It should be noted that we do not get physical solutions for $\phi_{\rm RCP}(S)$ for any value of $S$, rather only for $S >S_0$, where $S_0$ is a cutoff dependent on $\delta$. This is in agreement with earlier observations, where it has been shown that $S_0 = \delta - (1/\delta)$ \cite{smolalski2020identifying, meer2024bounded, supp}. We show this limit in Fig.~\ref{fig:deltaS}(b), marked by the dashed vertical lines. However, it should be noted that for a given value of skewness, there is an upper bound on the polydispersity given by $\delta_{\rm max} = (S+\sqrt{S^2+4})/2$, a result that we derived by solving the above quadratic equation in $\delta$, and which appears to be well satisfied in the calculations shown in Fig.~\ref{fig:deltaS}(a) (also check Supplementary Information) \cite{supp}. This is also the reason why the curves in Fig.~\ref{fig:deltaS} are not plotted for the entire range of $\delta$, but only for $\delta < \delta_{\rm max}$. The fact that there exists an upper limit on the achievable RCP density upon increasing the size polydispersity, and that this upper limit is controlled by the skewness, represents another important outcome of the current study.

 \begin{figure}[t]
    \centering
    \includegraphics[width=8.5cm]{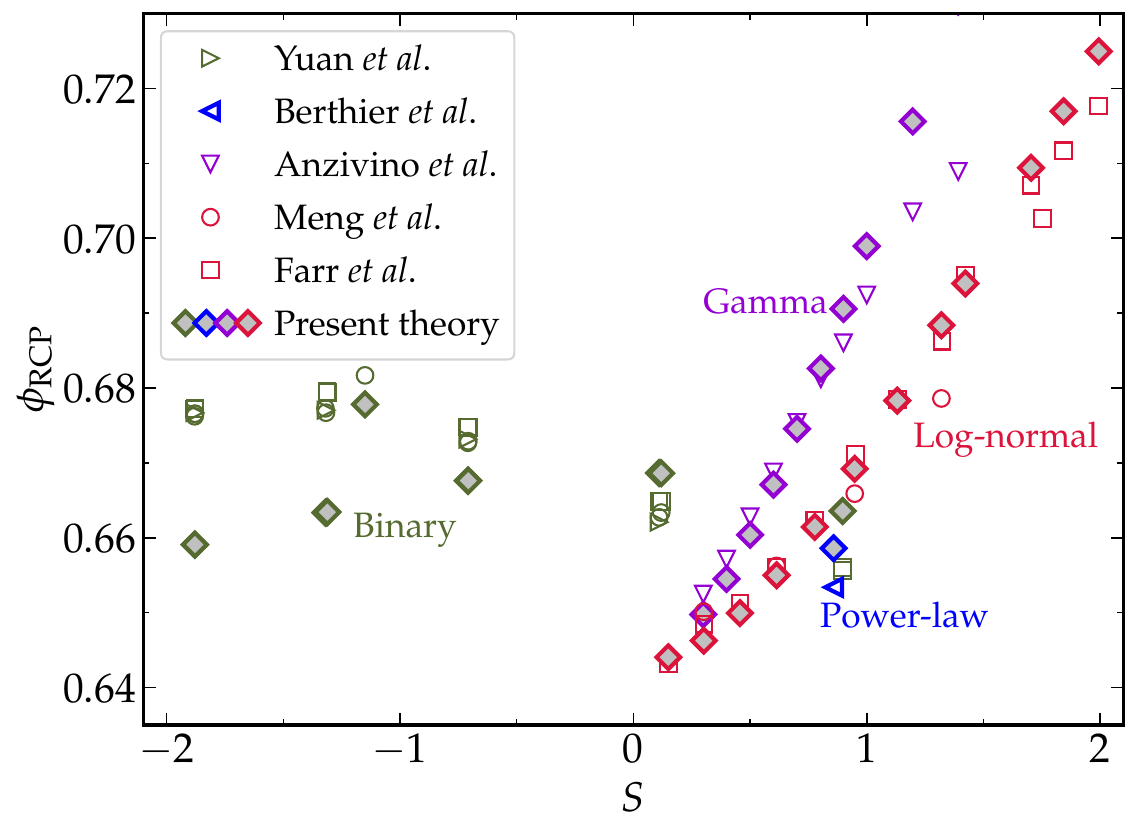}
    \caption{{ Comparison with simulation data.} $\phi_{\rm RCP}$ as a function of the skewness $S$ for binary mixtures (green), log-normal (red), power-law (blue), and gamma (violet) size distributions. Data has been adapted from Refs. \cite{yuan2018coupling} (Yuan {\it et al.}, open right triangles), \cite{berthier2016equilibrium} (Berthier {\it et al.}, open left triangle), \cite{anzivino2023estimating} (Anzivino {\it et al.}, open down triangles), \cite{meng2014packing} (Meng {\it et al.}, open circles), and \cite{farr2009close} (Farr {\it et al.}, open squares). These data correspond to a variety of $\delta$ values. The filled diamonds represent predictions from the present study, calculated for a given $S$ (as indicated on the horizontal axis) and the same $\delta$ value as the numerical data.}
    \label{fig:match}
\end{figure}

Now, we discuss the case of continuously polydisperse systems. We consider different diameter probability distributions $P(\sigma)$ defined for $\sigma > 0$. In particular, the following probability distributions are considered here, which have physical significance in different contexts: beta, generalized gamma, log-normal, and truncated power-law. Check the Supplementary Information \cite{supp} for the definition of these distribution functions, and the corresponding moments, coefficient of variation, and skewness. It should be noted that the generalized gamma distribution can be reduced, in special cases, to various other popular distributions, like Weibull, gamma, exponential, Rayleigh, half-normal, etc. 

The process to get the RCP volume fraction for such continuous size distributions is the same that we adopted for the binary mixture. Using the first three moments of the distribution, we solve Eq.~(\ref{polydisSol}) for $\phi_{\rm RCP}$ in terms of the shape parameter of the distribution function under consideration. Because it is not always possible to express the distribution parameters in terms of $\delta$ and $S$, we calculate $\phi_{\rm RCP}$ for different possible combinations of the parameters and at the same time the corresponding $\delta$ and $S$ values are recorded. In this way, it is possible to find the $\phi_{\rm RCP}$ for different combinations of $\delta$ and $S$. Similar to the case of binary mixtures, we can then identify the packing values for different $\delta$ values at given $S$, and for different $S$ values at given $\delta$.

In Figs.~\ref{fig:deltaS}(a) and (b), we present the RCP fraction values for four different continuous distributions as a function of $\delta$ for fixed $S$ and as a function of $S$ for fixed $\delta$ respectively, alongside the results for a binary mixture. Remarkably, the results for all distributions perfectly collapse onto the same master curve obtained for the binary mixture. The collapse onto master curves is a generic consequence of truncatable mixture equations of state, where thermodynamic properties depend only on a finite set of distribution moments \cite{sollich2002truncatable}. However, unlike the binary mixture, where a broad range of $S$ and $\delta$ can be explored (within the limits $S_0$ and $\delta_{\rm max}$, as discussed previously), the range of $\delta$ for a given $S$ or the range of $S$ for a given $\delta$ is bounded, and depends on the nature of the continuous distribution. In a previous work, Ogarko and Luding \cite{Luding} established that for any given polydisperse system an equivalent bidisperse system can be found which has the same equation of state in the fluid regime. Using numerical simulations, they also found that the polydisperse system and its bidisperse equivalent display similar jamming densities. See also Ref. \cite{Brouwers2024}. Our analytical findings thus align well with those previous numerical results. 

It is important to note that the polydispersity and skewness can either be independent or interdependent, depending on the probability distribution function. In other words, $\delta$ and $S$ can be either coupled or decoupled. For distributions with two shape parameters—such as the beta, generalized gamma, and truncated power-law distributions— $\delta$ and $S$ can be expressed independently of each other. Instead, for the log-normal distribution (which has only one shape parameter), this is not possible. Consequently, in Figs.~\ref{fig:deltaS}(a) and (b), the log-normal distribution yields only a single $\phi_{\rm RCP}$ value for a fixed $S$ or a fixed $\delta$, reflecting this inherent limitation.

Weeks and coworkers \cite{desmond2014influence, meer2024estimating} proposed that the RCP fraction follows an empirical relation of the form $\phi_{\rm RCP} = \phi_{\rm RCP}^{\rm M} + c(S-S_0)\delta^2$, where $\phi_{\rm RCP}^{\rm M}$ is the monodisperse random packing fraction and $c$ is a constant. Their analysis, based on fitting numerical data of $\phi_{\rm RCP}$ as a function of $S$, suggests that this empirical law holds for $\delta \leq 0.4$. However, as in Fig.~\ref{fig:deltaS}(b) shows, such a relation appears valid only for small $\delta$ values and in the low $S$ regime. For larger values of $\delta$ and in the large $S$ regime, $\phi_{\rm RCP}$ exhibits a nonlinear dependence, deviating from the proposed empirical law. Similar trends are also visible in Fig.~\ref{fig:deltaS}(a), where $\phi_{\rm RCP}$ as a function of $\delta$ deviates from $\phi_{\rm RCP} \sim \delta^2$ as $\delta$ increases.

Now, we compare the theoretical predictions of $\phi_{\rm RCP}$ from the present study for various size distributions with the numerical estimates from the literature. In Fig.~\ref{fig:match}, we show random close packing fraction values as a function of skewness for binary, gamma, log-normal, and power-law distributions (check figure caption for further details). The numerical data have been extracted from Refs.~\cite{yuan2018coupling} (binary distribution), \cite{berthier2016equilibrium} (power-law distribution), \cite{anzivino2023estimating} (gamma distribution), \cite{meng2014packing} (binary and log-normal distribution), and \cite{farr2009close} (binary and log-normal distribution). For each set of numerical data from the literature, the analytical calculations are based on solving Eq.~\eqref{polydisSol} with the same moments of the size distribution as in the numerical data. Hence, there are no free parameters in the comparison shown in Fig. \ref{fig:match}. 

In conclusion, we have developed fully analytical predictions for the effects of both skewness and polydispersity on the random close packing densities of polydisperse and bidisperse packings. Our analysis demonstrates that skewness, much like polydispersity, has a significant impact on the RCP density, which can be accurately calculated using our theoretical framework. Furthermore, we found that the RCP density as a function of skewness and polydispersity for several continuous size distributions collapses onto master curves obtained for bidisperse packings with the same $\delta$ and $S$, in agreement with prior computational studies \cite{desmond2014influence, meer2024estimating}. Importantly, our analytical theory reproduces previously reported numerical results for RCP density versus skewness across different distributions without any adjustable parameters. Overall, these findings deepen our understanding of the structure of disordered systems and establish a quantitative link between the shape of particle size distributions and jamming densities \cite{yuan2021connecting}. This connection has broad implications for various technological applications, including granular fluids \cite{petit2020additional, hara2021phase}, powder processing \cite{kumar2016tuning}, suspension rheology \cite{pednekar2018bidisperse}, biological systems \cite{schramma2023chloroplasts}, and even astronomical studies \cite{Blum}. Our results are fully consistent with earlier equilibrium-crowding approaches to polydisperse RCP \cite{anzivino2023estimating}. At the same time, the present theory provides a compact analytical treatment of skewness effects together with the correct exact asymptotic limits.

A.Z. and V.V. gratefully acknowledge funding from the European Union through Horizon Europe ERC Grant number: 101043968 “Multimech”. A.Z. also acknowledges financial support from the  US Army Research Office through contract nr. W911NF-22-2-0256, and from the Nieders{\"a}chsische Akademie der Wissenschaften zu G{\"o}ttingen in the frame of the Gauss Professorship program. 

\bibliography{references}
\end{document}